\documentclass[aps,prl,a4paper,twocolumn,showpacs,unsortedaddress,amsmath,amssymb]{revtex4}

\usepackage[latin1]{inputenc}
\usepackage{graphicx}
\usepackage{color}
\usepackage{dcolumn}
\usepackage{bm}
\usepackage{hyperref}
\usepackage{times}

\begin{document}

\title{Tuning the conductance of molecular junctions: transparent versus tunneling 
regimes}

\author{J. Ferrer}
\affiliation{Departamento de F\'{\i}sica, Universidad de Oviedo, 33007 Oviedo, Spain}

\author{V. M. Garc\'{\i}a-Su\'arez}
\affiliation{Department of Physics, Lancaster University, Lancaster LA1 4YB, U.K.}

\date{\today}

\begin{abstract}
We present a theoretical study of the transport characteristics of molecular junctions,
where first-row diatomic molecules are attached to (001) gold and platinum electrodes.
We find that the conductance of all of these junctions is of the order of the conductance
quantum unit $G_0$, spelling out that they belong to the transparent regime. We further find
that the transmission coefficients show wide plateaus as a function of the energy, instead
of the usual sharp resonances that signal the molecular levels in the tunneling regime.
We use Caroli's model to show that this is a rather generic property of the transparent regime
of a junction, which is driven by a strong effective coupling between the delocalized molecular
levels and the conduction channels at the electrodes. 
We analyse the transmission coefficients and chemical bonding of gold/Benzene and
gold/Benzene-dithiolate (BDT) junctions to understand why the later show large resistances,
while the former are highly conductive. 
\end{abstract}

\pacs{73.63.Rt,73.40.-c,73.63.-b}

\maketitle

\section{Introduction}
The field of molecular electronics was arisen by the early realization that organic molecules could 
act as rectifiers\cite{Rat74} when attached to conducting electrodes to form tunnel junctions. Many
experiments with a large variety of organic molecules have been
performed\cite{Tou,Col}, typically finding values of the conductance $G$ several orders of 
magnitude smaller than $G_0$ ($G_0=2\,e^2/h$ is the conductance quantum) and a large variability,
which hinder the reproducibility of the experiments. Molecular junctions can be understood in
terms of resonant tunneling models\cite{datta}, where the conduction is carried through the
Highest Occupied and Lowest Unoccupied Molecular Orbitals (HOMO and LUMO, respectively). 
These are revealed as sharp resonances in either the Densities of States
(DOS) of the molecule, or the Transmission coefficients ${\cal T}(E)$ of the junction, and are
usually located 1 or 2 eV above or below the Fermi level of the molecule, respectively. Conductance
values of the order of $G_0$ can only be achieved by pinning one of those resonances to the Fermi
level of the electrodes. Otherwise, the conductance is very low.

The invention of the Scanning Tunneling Microscope\cite{Binnig} allowed the fabrication of
stable atomic point contacts\cite{Gimzewski,Agr,Julio,jan,report}. These junctions were found
to be highly transparent in many cases, and to show values of the conductance of the order of
$G_0$, which confirmed early theoretical predictions on the matter\cite{Fer88,Fer90}. Theoretical
analyses of these junctions found that their transmission coefficients ${\cal T}(E)$ show wide
plateaus as a function of the energy $E$ of the incoming electrons, with heights of order one.
The high transparency of these
junctions is due to the good matching between the conduction channels at the electrodes and those
at the molecule. The (contact) resistance of the junction is different from zero because of the
different number of channels at the electrodes and the junction which leads to a
recombination of the former to match the later\cite{Fer88,datta}.

Importantly, more recent developments using MCBJ techniques demonstrated that high values of the
conductance ($\sim G_0$) are not restricted to atomic constrictions, but could also be obtained
even when platinum or palladium electrodes are bridged by hydrogen molecules\cite{Jan02,halxx}.
These results were reproduced by theoretical simulations\cite{Gar05,Thi}, which also showed how the
antibonding level of the hydrogen molecule hybridized strongly with the conduction channels of the
electrodes, providing a junction with a single conduction channel. The transparency of this channel
was manifested in the transmission coefficients ${\cal T}(E)$, that had a wide plateau of height
one. Furthermore, very recent experimental work confirms that junctions
comprising platinum electrodes and either simple benzene\cite{Jan08}, or a number of small 
molecules\cite{Jan08b} show conductance values of the order of $G_0$. 

As stated above, junctions in the tunneling regime can shed conductance values of order
$G_0$ provided that a molecular level is exactly pinned to the Fermi energy. But it is hard to
believe that all of the junctions discussed in those recent experiments\cite{Jan08,Jan08b}
display this pinning mechanism. Instead, they clearly indicate that highly
transparent molecular junctions can be fabricated with relative ease. They also indirectly hint 
that thiol-capping necessarily leads to junctions in the tunneling regime.

We have performed a number of transport simulations of molecular junctions where first-row
diatomic molecules are sandwiched by semi-infinite (001) gold and platinum electrodes, with our
code  SMEAGOL\cite{SME06}.
Our simulations confirm that this type of junctions is highly conductive. Indeed, the transmission
coefficients do not show resonant behavior but wide plateaus of height of order one instead. To
sustain theoretically
these simulations, we argue that the conductive behavior of a junction is determined by two 
factors. First, by the conjugation nature of the molecule, e.g.: whether the HOMO and LUMO levels
are delocalized throughout the whole molecule, or not. Second, by the strength of the chemical bond
between the conduction channels at the electrodes and these delocalized HOMO or LUMO orbitals. A
junction 
will be highly conductive provided that its molecule is conjugated and the chemical bond referred
above is strong.

\begin{figure}
\includegraphics[width=0.8\columnwidth,angle=-90]{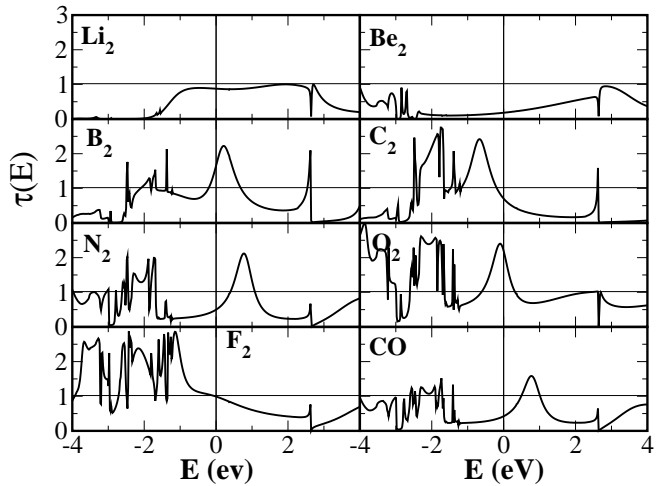}
\caption{Transmission coefficients at zero voltage as a function of energy 
${\cal T}(E)$, for a number of first-row diatomic molecules and CO attached to (001) gold
electrodes. The zero of energies corresponds to the position of the Fermi energy of the electrodes
at zero voltage. The distance between the last gold atoms at the electrodes and the molecules is set
to the equilibrium distance for the gold-$H_2$ junction (1.5 \AA).}
\end{figure}

Conjugated molecules are reasonably well approximated by Caroli's model\cite{Car}. We
use this model below to show that when the chemical bond between the delocalized HOMO or LUMO
orbitals and the
conduction channels is strong, the transmission coefficients show wide plateaus whose height is of
the order of $G_0$. We argue that these plateaus are robust against changes in the energy of the 
HOMO/LUMO levels, so that there
is no need to fine-tune their position to the Fermi level of the electrodes
in order to achieve large conductance values. On the contrary, when the conduction channels at the 
electrodes do not bind chemically with neither the HOMO nor the  LUMO levels then the junction is
in the tunneling regime where it shows resonant behavior.

Diatomic or triatomic molecules are sufficiently small to show conjugation.
Likewise, Benzene-based molecules are archetypical conjugated molecules, where the conjugation is 
driven by internal $\pi$-bonding. It is therefore
important to understand unequivocally why BDT junctions show strong tunneling behavior while
on the contrary the simpler Benzene junctions are highly transparent. We perform below a study of
the transmission coefficients of these junctions. Our study indicates that direct Au-S
bonding is detrimental of their conductive behavior.

\begin{figure}
\includegraphics[width=0.8\columnwidth,angle=-90]{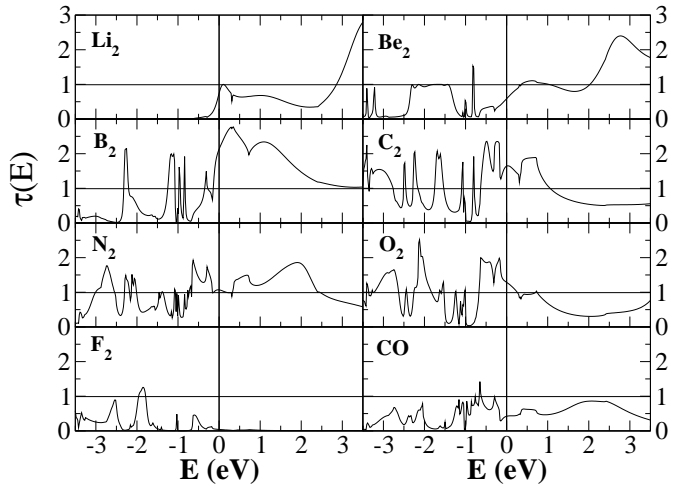}
\caption{Transmission coefficients at zero voltage as a function of
energy for a number of first-row diatomic molecules and CO attached to (001) platinum
electrodes. The geometry of the junctions has been relaxed in this case.}
\end{figure}

\section{Simulations of first-row diatomic molecules contacted to (001) gold and platinum
electrodes}
We have performed simulations of the conductance of (001) gold and platinum junctions that are
bridged by diatomic molecules of the first-row elements, using our code
SMEAGOL\cite{SME06}. SMEAGOL is a transport program which uses the Non Equilibrium Green's
Functions formalism\cite{datta} to compute the charge density and current of the junction.
The Hamiltonian of the junction is determined by DFT theory via the code SIESTA\cite{SIESTA}. 

In the case of gold junctions, we have taken for simplicity  a flat surface, oriented the molecules
perpendicular to it, and set the electrode-molecule distance
equal to the equilibrium distance of the gold-H$_2$ junction. We have used a more realistic geometry
for platinum in contrast, motivated by the recent experiments by Tal and coworkers\cite{Jan08b}. 
We have placed in these cases a pyramid of platinum atoms on top of the flat surfaces, as in our
earlier publications\cite{Gar05}. We have oriented the molecules in a variety of angles, including
the bridge, perpendicular and tilted orientations. We have finally relaxed the forces of the
atoms at the pyramids and molecules. We have found in this respect that Li$_2$, Be$_2$, B$_2$, 
C$_2$ and N$_2$ relax to the bridge position, while O$_2$, F$_2$ and CO relax to a tilted
orientation.

Our code computes the conductance of the junction via the formula
\begin{eqnarray}
G(V)&=&\frac{dI_{leads}(V)}{dV}\\&=&G_0\, \frac{d}{d(-eV)} \int\,d\omega\,T (E=\hbar
\omega,V)\,(n_L-n_R)\nonumber
\end{eqnarray}
\noindent where $T(E,V)$ are the energy- and voltage-dependent transmission coefficients of the
junction, defined in Eq. (17) in appendix A, and $n_{L,R}$ are the distribution functions of the
Left and Right electrodes, also defined in
Appendix A. $G(V)$ can be approximated at low enough voltages by the linear response formula
\begin{equation}
G(V)\simeq G_0 \,T(E=eV,0)=G_0\,{\cal T}(E=eV)
\end{equation}

We show in Figs. 1 and 2 the transmission coefficients ${\cal T}(E)$ for all the gold and 
platinum junctions simulated. Notice that we have taken the Fermi energy of the semi-infinite
electrodes at equilibrium as the reference energy. Overall, we find that the zero-voltage
conductance $G(0)$ is of order $G_0$, with the only exception of the
gold-$F_2$ junction. Further, all the transmission coefficients are rather smooth at positive
energies, while they show peaks in a range of energies below $E_F$.
These peaks do not correspond to molecular states, but rather to the d-band conduction channels at
the
electrodes, and are naturally located either a few $eV$ below $E_F$ for gold, or  extend up to $E_F$
for platinum. We will skip below any further reference to this d-band conduction channels, since
they are not relevant for our purposes.

A closer look at Fig. 1 shows that ${\cal T}$ for gold junctions is either a plateau
(as in Li$_2$, Be$_2$ and F$_2$), or the sum of a plateau and a broad resonance (in B$_2$, C$_2$,
N$_2$ O$_2$ and C). The low-voltage conductances are close to $G_0$ for Li$_2$ and
F$_2$, while they are of order 0.3-0.7 $G_0$ for Be$_2$, C$_2$, N$_2$ and CO, and of order 
1.5-2 $G_0$ for B$_2$ and O$_2$. 

The transmission coefficients of platinum junctions in Fig. 2 show a slightly
more complex structure, but are still pretty smooth and show no sign of the narrow
resonances that mark the tunneling regime. 
For platinum,  Li$_2$, and N$_2$ show conductances of about $G_0$, while Be$_2$ and CO have about
half a conductance quantum, and B$_2$, C$_2$ and O$_2$ have
conductances in the range 1.5-2 $G_0$. The conductance of F$_2$ is below 0.1 $G_0$. Our results
for CO agree with a previous simulation preformed by Strange and coworkers\cite{Str06}, but fail
to reproduce the peak at 1 $G_0$ shown in the experimental conductance histograms\cite{Jan08b}.
As a side remark, we should point out that we have found that the  conductance of platinum may
increase or decrease by even a factor of two depending on the placement and orientation of the
molecule.

\begin{figure}
\includegraphics[width=0.7\columnwidth,angle=-90]{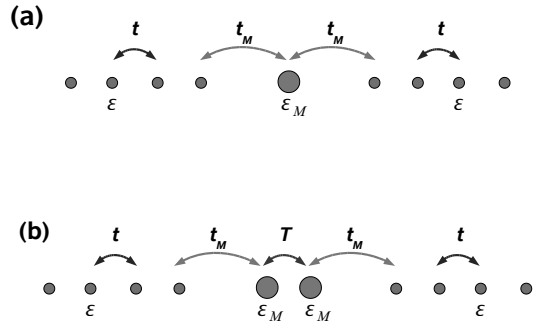}
\caption{(a) Two chains made up of $P$ atoms ($P\rightarrow\infty$),
connected to a central atom by a hopping matrix
element $t_M$. Each of the atoms in the chains has a single orbital with atomic energy
$\varepsilon$; electrons in the chains hop from one atom to the next via the hopping 
integrals $t$. The central atom has a single orbital of energy $\varepsilon_M$.
(b) Two atomic chains connected to a diatomic molecule. Each atom in the molecule has a single
orbital of energy $\varepsilon_M$; electrons hop between the two orbitals via the
intra-molecular hopping integral $T$. }
\end{figure}

\section{Analytical models}
Caroli's model in its simplest form, is depicted schematically in Fig. 3. It consists of
two identical semi-infinite chains, called Left (L) and Right (R) electrodes, 
which sandwich a free-standing atom (M), so that the three of them are initially unconnected.
The chains have a single orbital per atom of energy $\varepsilon$; electrons can hop between atoms
via the Hamiltonian matrix element $t$. The three pieces are initially held at the
same bias voltage, and the common Fermi energy $E_F$ of the unbiased chains is taken as the
reference energy. Since the chains are semi-infinite, they can be regarded as reservoirs of
electrons, subjected to a given equilibrium
chemical potential $\mu_{L,R}$, whose energy level population is described
by the Fermi distribution functions $n_{L,R}$. The central atom is held initially in
thermodynamic equilibrium  by its contact to a third reservoir at the arbitrary chemical potential 
$\mu_T=e\,V_T$. This determines the
equilibrium distribution of the atom $n_M$ and therefore its population. 
The whole system is subsequently biased by a voltage $V$ so that the chemical potentials
of the chains is shifted to $\mu_{L,R}=\pm e\,V/2$, while the chemical potential at the atom
can still be left equal to the initial value, although it is physically more reasonable to
reset it to the average between $\mu_L$ and $\mu_R$.
Notice that the three pieces stay initially in equilibrium since they are unconnected. 
Later on, the central atom is connected to the electrodes by adiabatically switching on the
hopping integrals between them until they reach their final value $t_M$. We assume that
the system is able to reach an stationary non-equilibrium state long time afterwards\cite{Dhar},
where there 
is a total bias voltage between the electrodes equal to $V$, that induces a finite electron current.
It is important to stress that the charge population at the atom is determined by its contact to
three reservoirs, each at a different chemical potential. Notice furthermore that, while the
coupling to the chains is controlled by $t_M$, the coupling to the third reservoir can not be
modulated, so that the flow of electrons between it and the atom is completely transparent.

This model can easily be solved analytically using the machinery of the Non-equilibrium Green's
functions formalism\cite{keldish,wagner} (see Appendix A for a thorough algebraic derivation). Fig.
4 shows the results for the Density of States at the atom (DOS)
\begin{equation}
\rho_M(\omega)= -\frac{1}{\pi}\,\mathrm{Imag}
\left[g_{MM}^R(\omega)\right]
\end{equation}
where  $g_{MM}^R$ is the retarded Green's function at the central atom, which is defined in appendix
A.
The figure also shows the conductance $G(V)$, which has been computed in this case performing
the numerical derivative of the current through the junction, to access the high-voltage behavior.

\begin{figure}
\includegraphics[width=0.8\columnwidth,angle=-90]{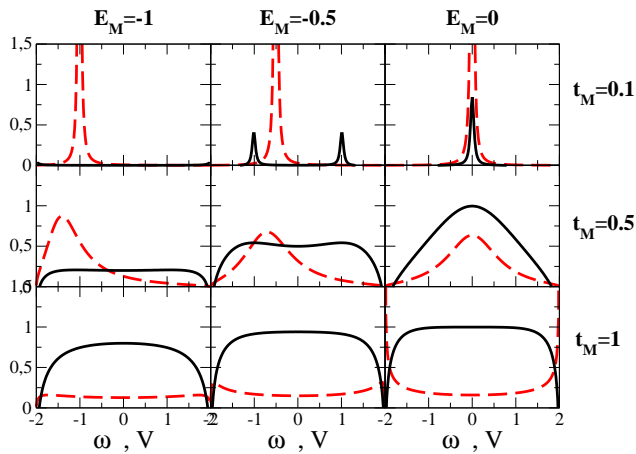}
\caption{(Color online) Conductance as a function of voltage (solid black curves), and density of
states
at the atom at zero voltage as a function of energy (dashed red curves), for Caroli's model. The
nine
panels displayed correspond to different choices of the parameters $t_M$, $\varepsilon_M$.}
\end{figure}

Notice that in this model the position of the HOMO/LUMO levels with respect to the Fermi energy 
of the electrodes (taken as 0) is given by the atomic energy $\varepsilon_M$, while the hopping
integral $t_M$ denotes the strength of the HOMO/LUMO hybridization with the electrodes. 
The hybridization causes two important and well-known effects on the bare atomic level
$\varepsilon_M$. First, 
it renormalizes it, e. g.: the level changes its energy. Second, its broadens it, since
now
an electron initially placed at the atomic level can hop back and forth to the electrodes, and 
hence the atomic state acquires a finite lifetime. These two effects are readily seen in $\rho_M$,
whereby the initial delta-like peak corresponding to
the atomic states moves and broadens to a resonance. The resonance is sharp if $t_M$ is much
smaller than $t$, or broad if $t_M$ and $t$ are of the same order of magnitude. What we wish to 
stress here is that the extent of this broadening also determines the behavior of the conductance 
of the junction. 

Fig. 4 illustrates how the different regimes of a junction manifest in the DOS and the conductance
and how these regimes are controlled by the basic parameters of the model. 
The top three panels correspond to the resonant tunneling regime. Notice that this regime is
seen even for values of the hybridization $t_M$ as large as 0.2. This regime is characterized by
a sharp resonance in the DOS, located at the energy position of the molecular orbital
$\varepsilon_M$. Additionally, sharp resonances appear in the conductance at different voltages,
whose maximum height can not exceed $G_0$. The low-voltage conductance is very small except if the
molecular orbital is pinned to the Fermi energy, in which case $G\simeq G_0$. We stress that
the molecular level must be fine-tuned to the Fermi energy to achieve sizable values of the
conductance. It is important to stress again that the resonant tunneling regime is 
achieved by values of $t_M$ which are smaller, {\em but not much smaller}, than $t$. Just a factor
of five is enough to place a junction within it.

The bottom panels in Fig. 4 correspond to highly transparent junctions, which corresponds to values
of $t_M\simeq 1$. The molecular orbital is in this case fully hybridized with the conduction
channels at the electrodes. The DOS is very broad, and has a width of the order of the
bandwidth of the conduction channels at the electrodes. More importantly, the conductance is very
flat and its height is very close to $G_0$. Notice that this happens for a wide range of positions
of the molecular orbital $\varepsilon_M$. Therefore, if the hybridization between the
molecular orbital and the conduction channels is large, then there is no need whatsoever to
fine-tune the position of the molecular orbital. In other words: when $t_M\simeq 1$, the molecular
orbital is always tuned, provided it is initially located within the band of the conduction
channels at the electrodes. This is one of the central results in this article.

The middle panels in Fig. 4 show the situation for junctions of intermediate transparency, The
DOS reflects this crossover behavior, where a resonance has already been developed but 
still has a large width. The conductance is very flat, but we point that its height has a 
stronger dependence on the position of the molecular level than was the case of the highly
transparent junction. 

\begin{figure}
\includegraphics[width=0.8\columnwidth,angle=-90]{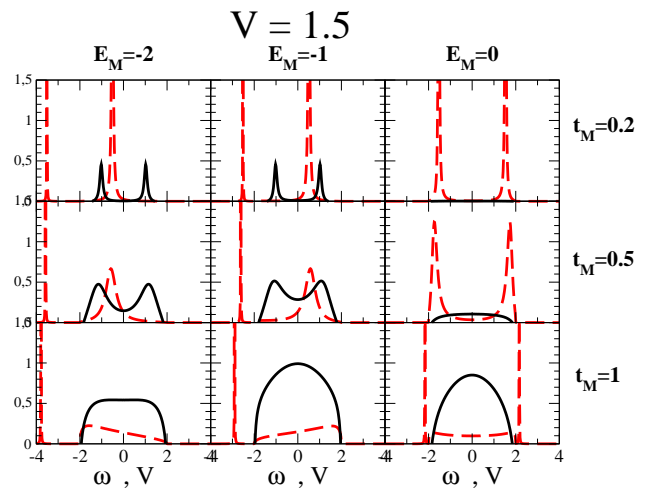}
\caption{(Color online)
Conductance as a function of voltage (solid black curves) and density of states
at the atom as a function of energy (dashed red curves) for the diatomic molecule model. The nine 
panels displayed correspond to different choices of the parameters $t_M$, $\varepsilon_M$. 
The parameter $T$ is set to 1.5 in all cases.}
\end{figure}

It is fortunate that some {\em Ab initio} simulation codes\cite{SIESTA} allow to make 
approximate estimates of the physical parameters that appear in Caroli's model. We have indeed
recently performed such estimates for a
constriction consisting of platinum electrodes bridged by $H_2$ molecules. We found that in this
case $t\sim t_M\sim 5\, eV$ (e.g.: $t_M/t\sim 1$), and that the anti-bonding state of
the molecule was located within the s-d-band complex of platinum\cite{Gar05}. We hence ascribed
that junction to the highly transparent regime, driven by the large value of $t_M$.
Most of the junctions that we have simulated in this article, on the other hand, correspond to the
crossover regime,
where the conductance is of order $G_0$, but where its exact value has a significant dependence on
the exact position of the molecular orbital. This is also the case of the Benzene junctions
discussed by Kiguchi and coworkers\cite{Jan08}. 

To make a closer contact with experiments on diatomic molecules, we have performed a slight
modification of Caroli's model. that we call the diatomic model. In it, the central atom is
replaced by a diatomic molecule as depicted in Fig. 1 (b). The nice feature of this model is 
that on the one hand, it accounts for both HOMO and LUMO levels and, on the other, it is also
easily solved analytically (the algebra is relegated to Appendix B).
This model can be applied directly to the case of the platinum-$H_2$ constriction, where we
showed that the electronic conduction is carried by the anti-bonding state of the molecule,
which is strongly hybridized to the platinum conduction bands, while 
the bonding state lies slightly below the edge of the platinum conduction band and hence does
not participate in the chemical bond, showing up as a sharp resonance in the DOS\cite{Gar05}.
As stated above, we estimated that for this case $t_M \sim t\sim 5 eV$ ($t_M/t\sim 1$), while 
$T$ was slightly larger, about 6 eV ($T/t\sim 1-1.5$).

We display in Fig. 5 the results of the diatomic model for the DOS at the
left atom in the molecule, and for $G(V)$ for the case where $T=1.5$, and where the bonding state
lies below the conduction band edge and hence shows up as a sharp resonance in the DOS. We again
find that when $t_M$ is much smaller
than $t$, both the DOS and $G$ display resonant behavior, whereby the conductance is always very
small, except for specific (usually too large) voltages, or when the anti-bonding state is finely
positioned at the Fermi energy. On the contrary, when $t_M\sim t$, the
resonance broadens, and the conductance has long plateaus with values close to $G_0$. 
The results that compare best to our simulation data for the DOS at the molecule in
the platinum-$H_2$ constriction (see Fig. 3 in Ref. [\onlinecite{Gar05}]) correspond to the
middle panel in the bottom row. Hence we expect that this model can describe correctly the
platinum-$H_2$ constriction if we use the parameters $T\sim1.5$, $t_M\sim1$ and 
$\varepsilon_M\sim -1$. Notice that the conductance indeed displays a value close to $G_0$.

A final note in this section relates to the large variability in the conductance obtained 
experimentally in BDT junctions. We remind that the voltage at which the conductance resonances 
depends strongly on the energy position of the molecular orbital for junctions belonging to the
resonant tunneling regime. In other words, the value of the measured conductance depends strongly on
the details of the orbital and its bonding for this type of junctions. On the contrary, since the 
conductance has a weak dependence on the details of the molecular orbital for highly transparent
junctions, we expect that the experimental variability must be suppressed for them.

\begin{figure}
\includegraphics[width=0.48\columnwidth]{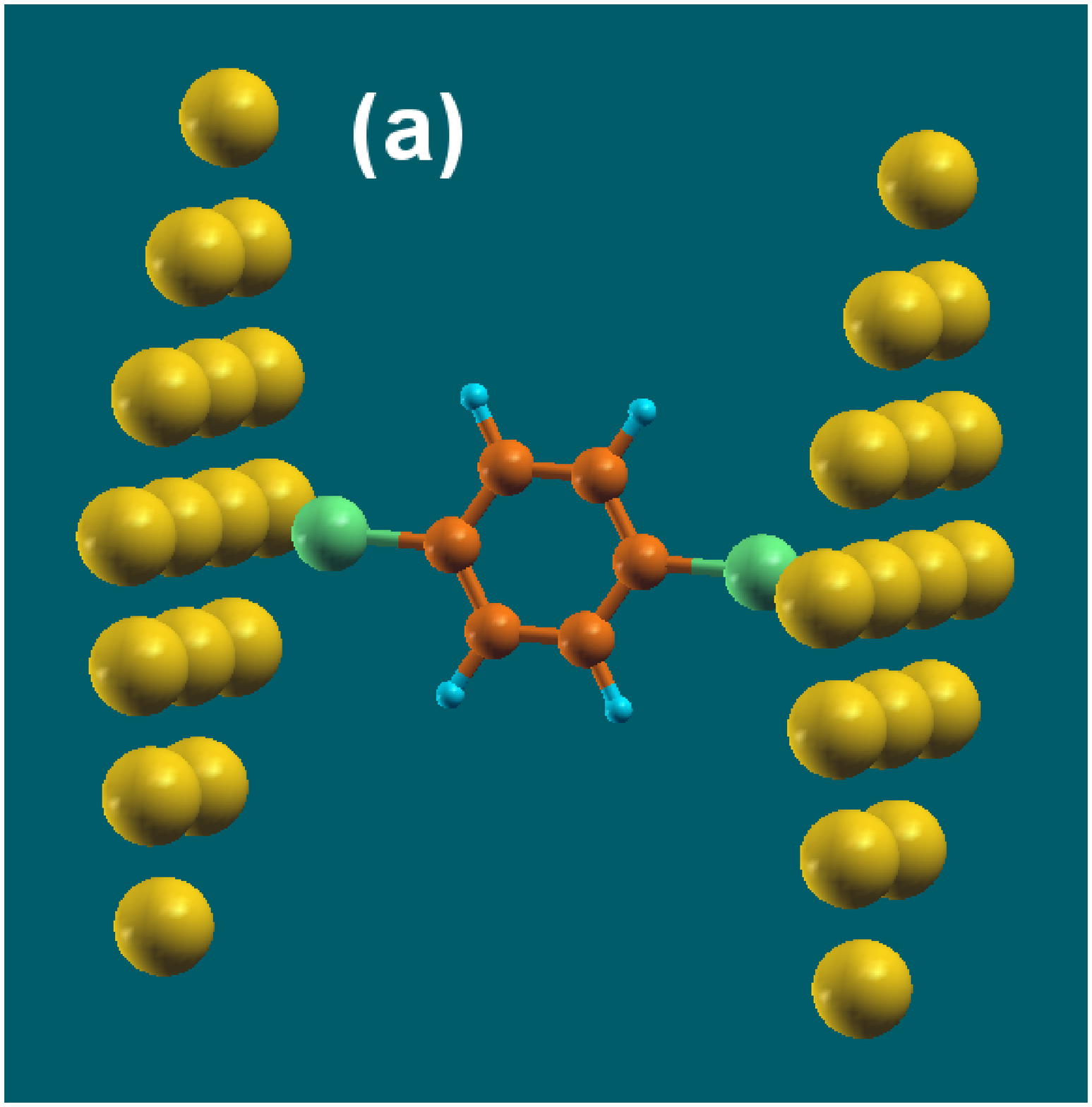}
\includegraphics[width=0.48\columnwidth]{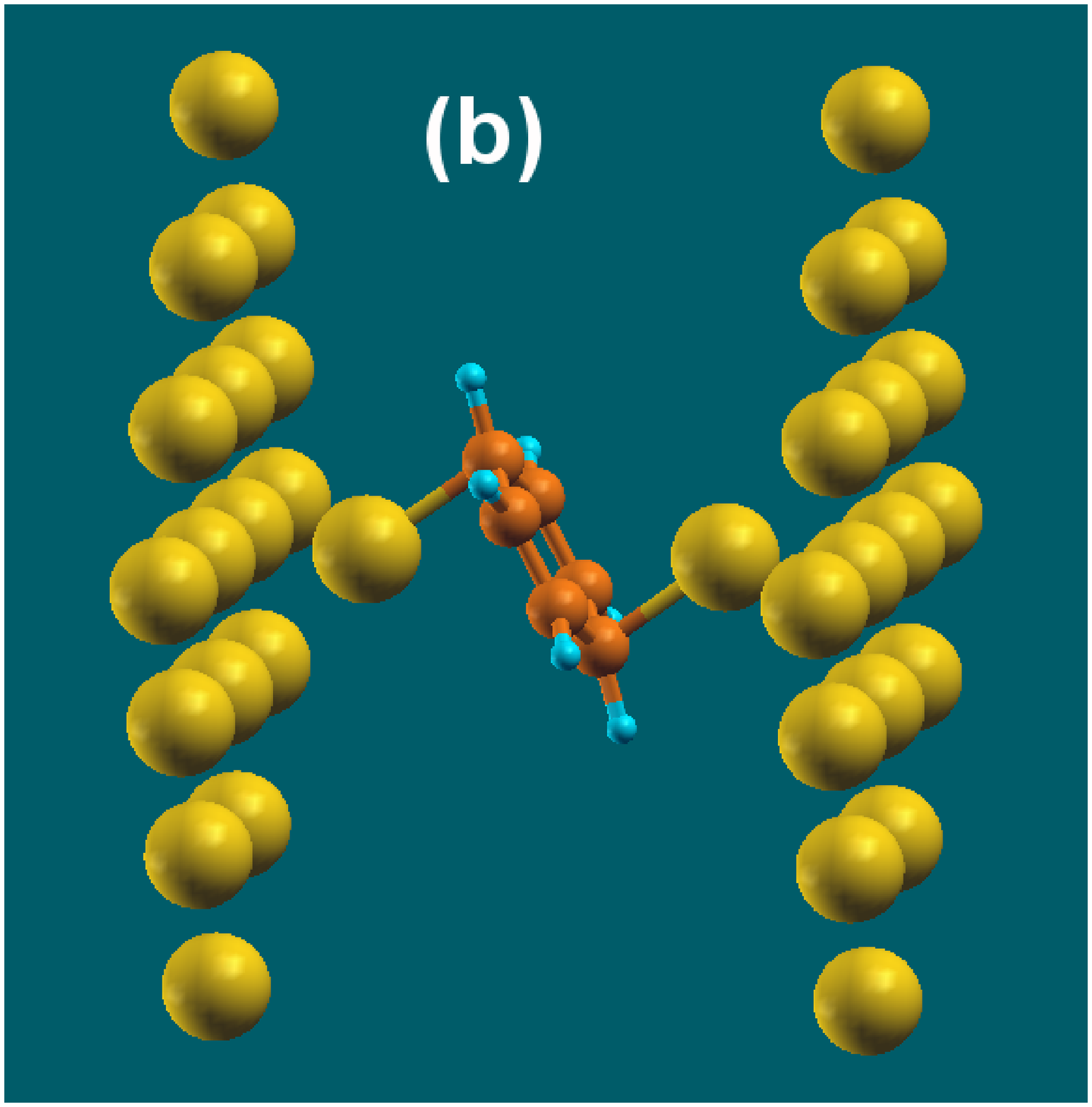}
\caption{(Color online) Geometry of (a) Benzene-dithiolate/gold, and  (b)
Benzene/gold junctions. The drawings show the last layer of gold atoms at each electrode and
the carbon, hydrogen and sulfur atoms of the molecules. Fig. (b) also shows the gold apex atoms 
attached to the gold flat surfaces.}
\end{figure}

\section{Analysis of Benzene and Benzene-dithiolate junctions}
Benzene and Benzene-dithiolate are conjugated molecules whose HOMO and LUMO states correspond to
$\pi$-bonded p orbitals, which are delocalized throughout the whole molecule.
BDT/gold structures are archetypical molecular junctions, which are used as
templates against which both theory and experiments are benchmarked. Experimental data show always
small
conductance values, which range from one to three orders of magnitude smaller than $G_0$, depending
on the experiment. Theoretical results tend to shed larger values for $G$, typically of the order
of
0.05 to 0.1 $G_0$. This discrepancy is usually attributed to a poor description of
electronic correlations by the theoretical models\cite{Sanvito}. BDT/gold junctions therefore seem
to belong
to the resonant tunneling regime described in the previous section. Interestingly, Benzene/platinum
junctions are highly conductive\cite{Jan08}, showing values of $G$ of the order of $G_0$, which
indicates that
these junctions belong to the transparent regime of Caroli's model. It is therefore important to 
understand better the contrasting behavior of these seemingly similar junctions.

In order to do so, we have simulated BDT junctions attached to (001) flat gold surfaces. 
BDT molecules attach via the thiol-groups, such that the sulfur atoms are placed 
in hollow sites at each side, as shown in Fig. 6 (a). We have set a sulfur/gold distance 
of 2 \AA, such that the junction is slightly stretched.
We have also simulated Benzene molecules attached to (001) gold surfaces. The surfaces are
initially flat, but we have placed an additional gold atom on top of each of them, to
provide for a preferred anchoring site. Benzene molecules bind to these electrodes via the
$\pi$-conjugated molecular orbitals, yielding a geometry for the junction such as that shown in 
Fig. 6 (b). We have set a distance of 5.0 \AA~ between the two apex gold atoms, which is again
slightly longer than the equilibrium distance of the junction. We note that there is no direct
hybridization between 
gold orbitals at each side of the junction for such a distance, which is confirmed by negligible 
transmission coefficients ${\cal T}(E)$.
Upon
relaxation of forces, we have found that the center of the Benzene plane is tilted in such a 
way to maximize the bonding between the $\pi$ orbitals and the apex gold atoms.

\begin{figure}
\includegraphics[width=0.75\columnwidth,angle=-90]{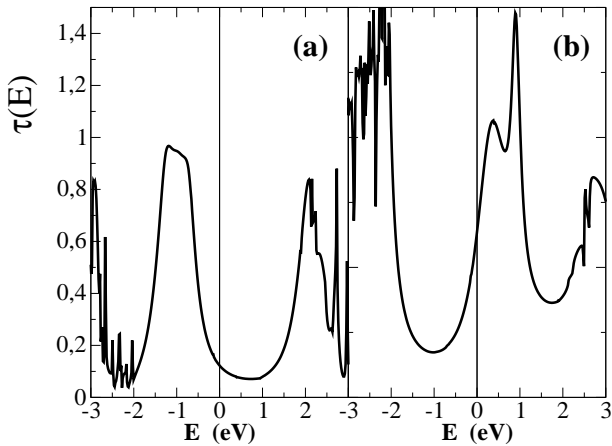}
\caption{Transmission coefficients ${\cal T}(E)$ for the two geometries in
the previous figure: a) BDT; (b) Benzene..}
\end{figure}

The transmission coefficients of both junctions are plotted in Figs. 7 (a) and (b). We find that
they both show broad resonances, which classifies them in a crossover ground between  the resonant
tunneling and intermediate transparency regimes. Consequently, the effective hybridization
between the HOMO/LUMO levels and the conduction channels is neither too large nor too small. We
therefore expect that the conductance of these junctions must have a relatively strong dependence
on the details of the junctions. BDT junctions have two main peaks peaks, placed 1 eV below and 2 eV
above the Fermi level, respectively. There is a gap in between, so that our simulations
yield a zero-voltage conductance of about 0.1 $G_0$. This height is actually reduced by strong
correlation effects\cite{Sanvito}.
The placement of the two peaks and the gap is due to the fact that the $\pi$-conjugated orbitals
are orthogonal to the
conduction channels of the electrodes, due to the geometry of the junction. 
The geometry of Benzene/gold junctions on the contrary is such that the $\pi$-conjugated orbitals
hybridize directly
with the conduction channels at the electrodes. Consequently, the broad resonance is moved, or 
pinned, to the Fermi level.

The reactivity of the thiol groups is therefore detrimental of the conductivity of BDT junctions.
In other words, BDT molecules attach to the electrodes via the thiol groups, yielding a junction
geometry where the 
conjugated orbitals do not bind to the conduction channels. When the thiol groups are missing,
Benzene attaches to the electrodes via the less reactive $\pi$ molecular orbitals. While the
$\pi$-gold chemical bond is less reactive, it has the virtue of providing a direct hybridization
between the HOMO/LUMO orbitals and the conduction channels at the electrodes.

\section{conclusions}
We have performed simulations of a number of molecular junctions, where (001) gold
or platinum electrodes sandwich first-row diatomic molecules. We have found that these
junctions are highly conductive, which is manifested both in large values of the conductance
and in smooth transmission coefficients ${\cal T}(E)$.  We have used Caroli's model to argue
that this is a generic feature of the transparent regime of a junction, which is driven by 
a high hybridization between the delocalized molecular orbitals in the molecule and the conduction
levels at the electrodes.

JF acknowledges conversations with J. van Ruitenbeek, as well as his sharing his results with us
prior to submission. This research has been funded by the Spanish government (project
FIS2006-12117).

\appendix
\section{Appendix A: Caroli's model}
It is convenient from a mathematical point of view to split the total Hamiltonian in
two pieces: the unperturbed Hamiltonian of the total system
${\cal H}_0$ and the perturbation that drives the system out of equilibrium ${\cal H}_1$.
We use the basis of atoms in the system, whose $2P+1$ states we denote by $|i>$. Then the 
Hamiltonians can be written as the following matrices 
\begin{equation}
{\cal H}_0=\left(\begin{array}{cccc}
\hat{H}_L&0&0\\
0& \varepsilon_{M} &0\\
0&0&\hat{H}_R
\end{array}\right)\,\,\,\,\,\,
{\cal H}_1=\left(\begin{array}{cccc}
0&\hat{T}_L&0\\
\hat{T}_{L}^{\dagger}& 0 &\hat{T}_{R}^{\dagger}\\
0&\hat{T}_{R}&0
\end{array}\right)\;.
\end{equation}
where $H_{L,R}$ are $P\times P$ tridiagonal matrices of the form
\begin{equation}
\hat{H}_{L,R}=\left(\begin{array}{cccccccc}
.&.&.& .              & .           &.&.&.\\
.&0&t&\varepsilon\pm eV/2& t           &0&.&.\\
.&.&0&t               & \varepsilon\pm eV/2 &t&0&.\\
.&.&.&.               &.            &.&.&.
\end{array}\right)\;.
\end{equation}
and 
 $\hat{T}_L^{\dagger}=(...,0,0,t_M)$ and $\hat{T}_R^{\dagger}=(t_M,0,0...)$ are 
$P$-dimensional vectors.

The retarded and advanced Green's functions of the unperturbed system can be calculated 
through the equations:
\begin{equation}
[\omega^{\pm}-{\cal H}_0]\,{\cal G}^{R,A}_0(w)={\cal I} 
\end{equation}
where $\omega^{\pm}=\omega\pm i \delta$, $\delta$ being an infinitesimal number.
Notice that ${\cal G}$ are also huge matrices of size $(2P+1)\times(2P+1)$.

This large set of coupled equations can actually be reduced enormously by gaussian
elimination 
of the atoms in the electrodes until only three states remain, ${|L>, |M>, |R>}$. The resulting 
3x3 matrix Green's functions are called $F$ to avoid confusing them
with the conductance, which is denoted by $G$. Their matrix elements (R/A superindices are
henceforth dropped when there is no danger of confussion)
\begin{equation}
 F_0 = \left(\begin{array}{ccc}
g_L & 0   & 0 \\
0   & g_M & 0 \\
0   & 0   & g_R
\end{array}\right)
\end{equation}
Here $g_{L,R}$ are the surface Green's function of the electrodes, and in our one-dimensional
model they are equal to
\begin{equation}
g_{L,R}^{R,A}= \frac{2}{\omega^{\pm}-\varepsilon+\sqrt{(\omega^{\pm}-\varepsilon)^2-4 t^2}}
\end{equation}
\noindent To simplify matters, we will take henceforth $\varepsilon=0$ and $t$ as the energy
unit. We also define a {\em lesser} Green's function, that carries information of the
electron occupation in each piece,
\begin{equation}
 F_0^< = \left(\begin{array}{ccc}
g_L^< & 0   & 0 \\
0   & g_M^< & 0 \\
0   & 0   & g_R^<
\end{array}\right)
\end{equation}
where each $g^<_{L,R,M}(\omega)$ is written in terms of the density of
states $\rho_{L,M,R}$ and the Fermi distribution function $n_{L,M,R}$ of each unconnected piece, as
$g_i^<(\omega)=2\,\pi\,i\,\rho_i(\omega-e\,V_i)\,n(\omega-e\,V_i)$.

Then the Keldish formalism\cite{keldish,Car} provides a simple recipe to compute the Green's
functions of the final system in the steady state,
\begin{equation}
\begin{array}{cc}
&F^R=[[G_0^R]^{-1}-{\cal H}_1]^{-1}=
\left(\begin{array}{ccc}
g_{LL}^R & g_{LM}^R   & g_{LR}^R \\
g_{ML}^R   & g_{MM}^R & g_{MR}^R \\
g_{RL}^R   & g_{RM}^R   & g_{RR}^R
\end{array}\right)
\\\\
&F^<=G^R\,[G^R_0]^{-1}\,G^<_0\,[G^A_0]^{-1}\,G^A= \left(\begin{array}{ccc}
g_{LL}^< & g_{LM}^<   & g_{LR}^< \\
g_{ML}^< & g_{MM}^<   & g_{MR}^< \\
g_{RL}^< & g_{RM}^<   & g_{RR}^<
\end{array}\right)
\end{array}
\end{equation}
These equations can easily be solved analytically, with the following result
for the retarded/advanced Green's function,
\begin{equation}
t_M^2\,F=\frac{1}{D}
\left(\begin{array}{ccc}
\Sigma_L\,(\omega-\varepsilon_M-\Sigma_R) & t_M\,\Sigma_L  & \Sigma_L\,\Sigma_R  \\
t_M\,\Sigma_L                             & t_M^2        & t_M\,\Sigma_R         \\
\Sigma_L\,\Sigma_R                      & t_M\,\Sigma_R & \Sigma_R\,(\omega-\varepsilon_M-\Sigma_L)
\end{array}\right)
\end{equation}
where $\Sigma_{L,R}=t_M^2\, g_{L,R}$ and $D=\omega-\varepsilon_M-\Sigma_L-\Sigma_R$.
Likewise, $F^<$ can be written as the sum of the following three matrices
\begin{widetext}
\begin{equation*}
L=\frac{2\,\pi\,i\,\rho_L\,n_L}{|D|^2}
\left(\begin{array}{ccc}
|\omega-\varepsilon_M-\Sigma_R|^2 & t_M\,(\omega-\varepsilon_M-\Sigma_R^R)  & 
\Sigma_R^A\,(\omega-\varepsilon_M-\Sigma_R^R)  \\
t_M\,(\omega-\varepsilon_M-\Sigma_R^A) & t_M^2        & t_M\,\Sigma_R^A         \\
\Sigma_R^R\,(\omega-\varepsilon_M-\Sigma_R^A)  & t_M\,\Sigma_R^R  &|\Sigma_R|^2
\end{array}\right)
\end{equation*}
\begin{equation}
R=\frac{2\,\pi\,i\,\rho_R\,n_R}{|D|^2}
\left(\begin{array}{ccc}
|\Sigma_L|^2 & t_M\,\Sigma_L^R  & \Sigma_L^R\,(\omega-\varepsilon_M-\Sigma_L^A) \\
t_M\,\Sigma_L^A  & t_M^2        & t_M\,(\omega-\varepsilon_M-\Sigma_L^A)  \\
\Sigma_L^A\,(\omega-\varepsilon_M-\Sigma_L^R) & t_M\,(\omega-\varepsilon_M-\Sigma_L^R)  &
|\omega-\varepsilon_M-\Sigma_L|^2
\end{array}\right)
\end{equation}
\begin{equation*}
t_M^2\,M=\frac{2\,\pi\,i\,\rho_M\,n_M}{|g_M\,D|^2}
\left(\begin{array}{ccc}
|\Sigma_L|^2 & t_M\,\Sigma_L^R  & \Sigma_L^R\,\Sigma_R^A \\
t_M\,\Sigma_L^A     & t_M^2           & t_M\,\Sigma_R^A          \\
\Sigma_R^R\,\Sigma_L^A & t_M\,\Sigma_R^A  & |\Sigma_R|^2
\end{array}\right)
\end{equation*}
\end{widetext}
These lesser Green's functions enter the calculation of the electronic charge and current. For
instance, the charge at the atom $N_M$ can be written as follows:
\begin{eqnarray}
N_M&=&\int\frac{d\omega}{2\,\pi\,i}\,g_{MM}^<(\omega)\\
&=&\int \frac{d\omega}{2\,\pi} \,
|g_{MM}^R|^2 \,\left(\Gamma_L\,n_L+\Gamma_R\,n_R +
\frac{\Gamma_M}{|t_M\,g_M|^2}\,n_M\right)\nonumber
\end{eqnarray}
This formula for the charge can be rewritten in terms of the conventional
expression for the equilibrium state
\begin{equation}
N_{eq}=
-\int \frac{d\omega}{\pi}\,\mathrm{Im}\,\left[g_{MM}^R\right]\,n_L
\end{equation}
plus an explicit non-equilibrium term
\begin{eqnarray}
N_{non-eq}=&&\int \frac{d\omega}{2\,\pi} \,
|g_{MM}^R|^2\,\Gamma_R\,(n_R-n_L)\nonumber\\&+&\int \frac{d\omega}{2\,\pi} \,
|g_{MM}^R|^2\,\frac{\Gamma_R}{|t_M\,g_M|^2}\,(n_M-n_L)
\end{eqnarray}
\noindent where
$\Gamma_{L,R}=i\,(\Sigma_{L,R}^R-\Sigma_{L,R}^A)=2\,\pi\,t_M^2\,\rho_{L,R}(\omega-eV_{L,R})$, 
and likewise $\Gamma_{M}=2\,\pi\,t_M^2\,\rho_M(\omega-eV_{T})$. 

The current traversing the left contact is computed via the formula
\begin{equation}
I_{LM}= -\frac{2\,e\,t_M}{\hbar}\,\int \frac{d\omega}{2\,\pi}\,
\left(g_{LM}^<-g_{ML}^<\right)
\end{equation}
which can be written as a sum of the following two contributions,
\begin{eqnarray}
I_{leads}&=&-\frac{G_0}{e}\,\int\,d\omega\,|g_{MM}^R|^2\,\Gamma_L\,\Gamma_R\,(n_L-n_R)
\nonumber\\&=&
-\frac{G_0}{e}\,\int\,d\omega\,T(E=\hbar \omega,V)\,(n_L-n_R)
\nonumber\\
I_{atom,L}&=&-\frac{G_0}{e}\,\int\,d\omega\,|g_{MM}^R|^2\,\frac{\Gamma_L\,\Gamma_M}{t_M^2}\,n_M
\end{eqnarray}
\noindent Notice that the current traversing the right contact can be written as the sum of
$I_{leads}$ plus $I_{atom,R}$ where 
\begin{equation}
 I_{atom,R}=-\frac{G_0}{e}\,\int\,d\omega\,|g_{MM}^R|^2\,\frac{\Gamma_R\,\Gamma_M}{t_M^2}\,n_M
\end{equation}
\noindent Since these atomic contributions apparently break the conservation of charge,
$I_{atom,(L,R)}$ are usually dropped (as is the second term in Eq. (15)). In other words, the 
total current $I$ is approximated by $I_{leads}$.

\section{Appendix B: Diatomic molecule model}
The unperturbed Hamiltonian and the perturbation in the diatomic model take the form:
\begin{equation}
{\cal H}_0=\left(\begin{array}{ccccc}
\hat{H}_L&0&0\\
0& \varepsilon_{M} &T&0\\
0&T&\varepsilon_M&0\\
0&0&0&\hat{H}_R
\end{array}\right)\;.
\end{equation}
where $H_{L,R}$ are the $P\times P$ tridiagonal matrices described in Appendix A, and 
\begin{equation} 
{\cal H}_1=\left(\begin{array}{ccccc}
0&\hat{T}_L&0&0\\
\hat{T}_{L}^{\dagger}& 0 & 0 &0\\
0                    & 0 &\hat{T}_{R}^{\dagger}&0\\
0&0&\hat{T}_{R}&0
\end{array}\right)\;.
\end{equation}
After decimating the unwanted degrees of freedom, the unpertubed Green's functions look like
\begin{equation}
 F_0^{R,A,<} = \left(\begin{array}{ccc}
g_L^{R,A,<} & 0   & 0 \\
0   & \hat{g}_M^{R,A,<} & 0 \\
0   & 0   & g_R^{R,A,<}
\end{array}\right)
\end{equation}
where 
\begin{equation}
\hat{g}_M^{R,A}=\left(
\begin{array}{cc}
\omega-\varepsilon_M	&	-T	\\
-T			&	\omega-\varepsilon_M
\end{array}
\right)
\end{equation}
and $\hat{g}_M^<=-1/\pi\,(\hat{g}_M^R-\hat{g}_M^A)\,n_M$.
The algebra is more tedious, but the final result for the retarded Green's function 
$t_M^2\,F$  is the following:
\begin{widetext}
\begin{equation}
\frac{1}{D}
\left(\begin{array}{ccccc}
\Sigma_L\,[(\omega-\varepsilon_M)(\omega-\varepsilon_M-\Sigma_R)-T^2] & 
t_M\,\Sigma_L\, (\omega-\varepsilon_M-\Sigma_R) & t_M\,T\,\Sigma_L &T\,\Sigma_L\,\Sigma_R \\
t_M\,\Sigma_L\,(\omega-\varepsilon_M-\Sigma_R)   & t_M^2\,(\omega-\varepsilon_M-\Sigma_R)   &
t_M^2\,T&t_M\,T\,\Sigma_R      \\
t_M\,T\,\Sigma_L   &  t_M^2\,T  & t_M^2\,(\omega-\varepsilon_M-\Sigma_L) &
t_M\,\Sigma_R\,(\omega-\varepsilon_M-\Sigma_L)  \\
T\,\Sigma_L\,\Sigma_R                      & t_M\,T\,\Sigma_R &
t_M\,\Sigma_R\,(\omega-\varepsilon_M-\Sigma_L) &
\Sigma_R\,[(\omega-\varepsilon_M)(\omega-\varepsilon_M-\Sigma_L)-T^2]
\end{array}\right)
\end{equation}
\end{widetext}
where $D=(\omega-\varepsilon_M)(\omega-\varepsilon_M-\Sigma_R-\Sigma_R)-T^2+\Sigma_L\,\Sigma_R$,
and the equation for the current traversing the left link can be written as
\begin{equation}
I_{LM}= -\frac{2\,e}{h}\int\,d\omega\, \Gamma_L\,\Gamma_R\,|G_{MM'}^R|^2\,(n_r-n_L)
\end{equation}
where the atomic contribution has been neglected again, and $G_{MM'}$ refers to the
element
(2,3) of the retarded Green's function matrix, that connects the two atoms in the molecule.

\end{document}